\documentclass[conference]{IEEEtran}
\IEEEoverridecommandlockouts
\usepackage{cite}
\usepackage{amsmath,amssymb,amsfonts}
\usepackage{graphicx}
\usepackage{subcaption}
\usepackage{textcomp}
\usepackage{xcolor}
\usepackage{url}
\usepackage{makecell}
\usepackage{mathtools}
\usepackage{float}
\usepackage{gensymb}
\usepackage{booktabs}
\usepackage{bm}
\usepackage{algorithm}
\usepackage{algpseudocode}
\usepackage[export]{adjustbox}

\def\BibTeX{{\rm B\kern-.05em{\sc i\kern-.025em b}\kern-.08em T\kern-.1667em\lower.7ex\hbox{E}\kern-.125emX}}

\usepackage[bottom=1.04in,top=0.75in,left=0.63in,right=0.63in]{geometry}
\begin{document}

\bstctlcite{IEEEexample:BSTcontrol}


\title{On the Spectral Efficiency of Indoor Wireless Networks with a Rotary Uniform Linear Array}

\author{Eduardo Noboro Tominaga\IEEEauthorrefmark{1}, Onel Luis Alcaraz López\IEEEauthorrefmark{1}, Tommy Svensson\IEEEauthorrefmark{2}, Richard Demo Souza\IEEEauthorrefmark{3}, Hirley Alves\IEEEauthorrefmark{1}\\

	\IEEEauthorblockA{
		\IEEEauthorrefmark{1}6G Flagship, Centre for Wireless Communications (CWC), University of Oulu, Finland.\\        
		E-mail: \{eduardo.noborotominaga, onel.alcarazlopez, hirley.alves\}@oulu.fi\\
        \IEEEauthorrefmark{2}Department of Electrical Engineering, Chalmers University of Technology, 412 96 Gothenburg, Sweden.\\ E-mail: tommy.svensson@chalmers.se\\
		\IEEEauthorrefmark{3}Department of Electrical and Electronics Engineering, Federal University of Santa Catarina (UFSC), Florian\'{o}polis,\\88040-370, Brazil. E-mail: richard.demo@ufsc.br\\
	}
}

\maketitle

\begin{abstract}
Contemporary wireless communication systems rely on Multi-User Multiple-Input Multiple-Output (MU-MIMO) techniques. In such systems, each Access Point (AP) is equipped with multiple antenna elements and serves multiple devices simultaneously. Notably, traditional systems utilize fixed antennas, i.e., antennas without any movement capabilities, while the idea of movable antennas has recently gained traction among the research community. By moving in a confined region, movable antennas are able to exploit the wireless channel variation in the continuous domain. This additional degree of freedom may enhance the quality of the wireless links, and consequently the communication performance. However, movable antennas for MU-MIMO proposed in the literature are complex, bulky, expensive and present a high power consumption. In this paper, we propose an alternative to such systems that has lower complexity and lower cost. More specifically, we propose the incorporation of rotation capabilities to APs equipped with Uniform Linear Arrays (ULAs) of antennas. We consider the uplink of an indoor scenario where the AP serves multiple devices simultaneously. The optimal rotation of the ULA is computed based on estimates of the positions of the active devices and aiming at maximizing the per-user mean achievable Spectral Efficiency (SE). Adopting a spatially correlated Rician channel model, our numerical results show that the rotation capabilities of the AP can bring substantial improvements in the SE in scenarios where the line-of-sight component of the channel vectors is strong. Moreover, our proposed system is robust against imperfect positioning estimates.
\end{abstract}


\section{Introduction}

\par Multi-User Multiple-Input Multiple-Output (MU-MIMO) technologies are instrumental in enhancing the performance of wireless communication networks. In MU-MIMO networks, a base station or Access Point (AP) equipped with multiple antenna elements serves multiple devices simultaneously. Relying on beamforming techniques, MIMO offers several benefits such as diversity and array gains, spatial multiplexing and interference suppression capabilities that collectively enhance the capacity, reliability and coverage of wireless systems \cite{heath2018}.


\par Traditional wireless communication systems adopt static antennas, i.e. antennas without any active movement capabilities. Nevertheless, the idea of antennas with movement capabilities has recently become popular among the research community. Movable antennas are able to exploit the wireless channel variation in the continuous spatial domain. This additional degree of freedom can enhance the quality of the wireless links and consequently the communication performance.

\par The performance of MU-MIMO systems with movable antennas has been studied in many recent works, e.g. \cite{zhu2023_1,zhu2023_2,xiao2023,pi2023,ma2023}. Nevertheless, their proposed movable antenna systems, which allow the antennas to move in a confined 2D or 3D region, require many components such as mechanical controllers, drivers and cables. Thus, they are complex, bulky, expensive to be deployed and maintained, present a high power consumption, and might present mechanical robustness issues. Rotary antennas, which are antennas or antenna arrays equipped with a high speed, high precision servo motor that controls its angular position, are a simpler alternative. Indeed, rotary antenna systems present lower complexity, cost, form factor and power consumption than the systems proposed in \cite{zhu2023_1,zhu2023_2,xiao2023,pi2023,ma2023}. The performance of a point-to-point Line-of-Sight (LoS) MIMO system where both the transmitter and receiver are equipped with Rotary Uniform Linear Arrays (RULAs) of antennas was studied in \cite{lozano2021}. This work showed that the RULAs can be configured to closely approach the LoS channel capacity. López \textit{et. al.} \cite{onel2022} and Lin \textit{et. al.} \cite{lin2023} studied the utilization of RULAs for wireless energy transfer. In their scenario, a power beacon equipped with a RULA is constantly rotating and transmitting an energy signal in the downlink. Multiple low-power devices harvest the energy from the transmitted signal to charge their batteries.


\par In this work, we propose an implementation of an indoor wireless network where the serving AP is equipped with a RULA. Differently than \cite{lozano2021}, we consider a MU-MIMO scenario, i.e., the AP equipped with a RULA serves multiple single-antenna devices simultaneously. Moreover, differently from \cite{onel2022,lin2023}, we evaluate an uplink data transmission scenario where the angular position of the RULA is set according to the position of the devices. Adopting a spatially correlated Rician channel model, we compare different linear receive combining schemes in terms of per-user mean achievable Spectral Efficiency (SE). Our numerical results show that the optimal rotation of the RULA brings substantial SE performance gains when the LoS components of the channel vectors are strong, even if the quality of the positioning information of the system is poor.

\begin{figure*}
    \centering
    \begin{minipage}[t]{0.475\textwidth}
        \centering
        \includegraphics[scale=0.4]{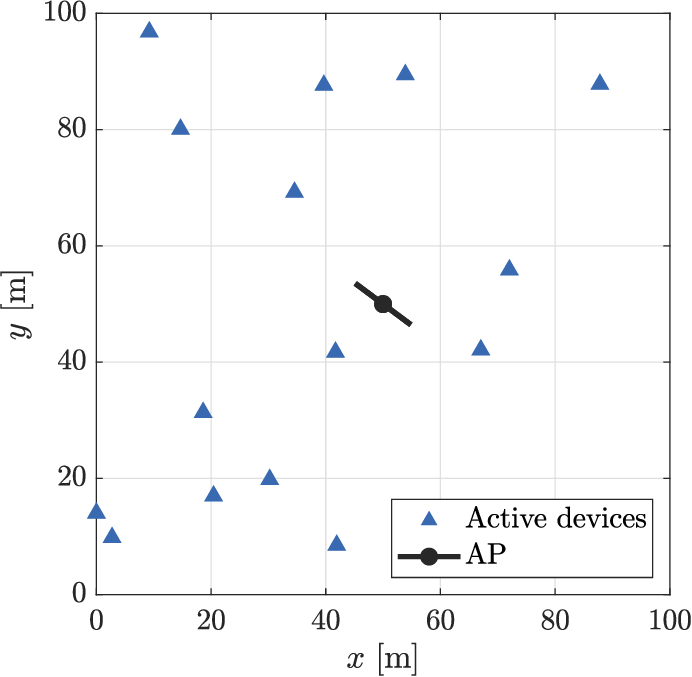}
        \caption{Illustration of the considered system model for $l=100$~m and $K=16$.}
        \label{illustrationScenario}
    \end{minipage}
    \hspace{0.4cm}
    \begin{minipage}[t]{0.475\textwidth}
        \centering
        \includegraphics[scale=0.6]{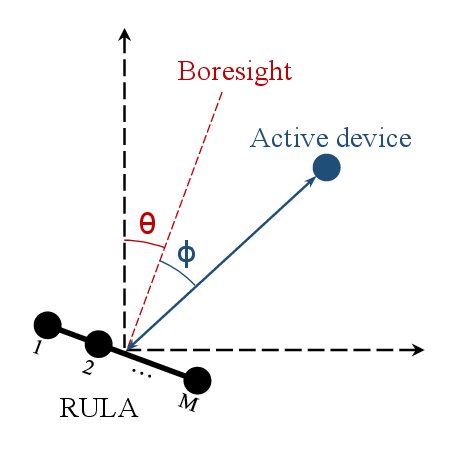}
        \caption{Illustration of a RULA, its distance and its angular position with respect to an active device.}
        \label{illustrationRULA}
    \end{minipage}
\end{figure*}


\par This paper is organized as follows. The considered system model is introduced in Section \ref{systemModel}. The mechanism for the optimization of the angular position on the RULA based on position estimates is presented in Section \ref{optimalAngularPosition}. A mathematical model for the positioning error is proposed in Section \ref{positioningErrorModel}. Numerical results and discussions are presented in Section \ref{numericalResults}. Finally, we draw the conclusions of this work in Section \ref{conclusions}.

\par \textbf{Notation:} lowercase bold face letters denote column vectors, while boldface upper case letters denote matrices. $a_i$ is the $i$-th element of the column vector $\textbf{a}$, while $\textbf{a}_i$ is the $i$-th column of the matrix $\textbf{A}$.  $[A]_{i,j}$ is the $i$-th row, $j$-th column element of the matrix $\textbf{A}$.
$\textbf{I}_M$ is the identity matrix with size $M\times M$. The superscripts $(\cdot)^T$ and $(\cdot)^H$ denote the transpose and the conjugate transpose operators, respectively. The magnitude of a scalar quantity or the cardinality of a set are denoted by $|\cdot|$. The Euclidean norm of a vector (2-norm) is denoted by $\Vert\cdot\rVert$. We denote the one dimensional uniform distribution with bounds $a$ and $b$ by $\mathcal{U}(a,b)$. We denote the multivariate Gaussian distribution with mean $\mathbf{a}$ and covariance $\mathbf{B}$ by $\mathcal{N}(\mathbf{a},\mathbf{B})$.

\section{System Model}
\label{systemModel}

\par We consider a square area with dimensions $l\times l\;\text{m}^2$, wherein $K$ single-antenna devices are served by a single AP equipped with a RULA. The AP is placed at height $h_{\text{AP}}$ and is equipped with $M$ half-wavelength spaced antenna elements. 
Let $\textbf{p}_k=(x_k,y_k)^T$ denote the coordinates of the $k$-th device, assuming for simplicity that all devices are positioned at the same height $h_{\text{device}}$. The system model is illustrated in Fig. \ref{illustrationScenario} and we focus on the uplink. Moreover, a RULA and its angular position with respect to a device are illustrated in Fig. \ref{illustrationRULA}, where $\theta\in[0,\pi]$ denotes the rotation performed by the servo-motor.

\subsection{Channel Model}

\par We adopt a spatially correlated Rician fading channel model \cite{ozdogan2019}. Let $\textbf{h}_k\in\mathbb{C}^{M\times1}$ denote channel vector between the $k$-th device and the AP. It can be modeled as \cite{dileep2021}
\begin{equation}
    \label{channelVector}
    \textbf{h}_k=\sqrt{\dfrac{\kappa}{1+\kappa}}\textbf{h}_k^{\text{los}} + \sqrt{\dfrac{1}{1+\kappa}}\textbf{h}_k^{\text{nlos}},
\end{equation}
where $\kappa$ is the Rician factor, $\textbf{h}_k^{\text{los}}\in\mathbb{C}^{M\times1}$ is the deterministic LoS component, and $\textbf{h}_k^{\text{nlos}}\in\mathbb{C}^{M\times1}$ is the random NLoS component.

\par The deterministic LoS component is given by
\begin{equation}
    \label{losComponent}
    \textbf{h}_k^{\text{los}}=\sqrt{\beta_k}
    \begin{bmatrix}
        1\\
        \exp(-j2\pi\Delta\sin(\phi_k))\\
        \exp(-j4\pi\Delta\sin(\phi_k))\\
        \vdots\\
        \exp(-j2\pi(S-1)\Delta\sin(\phi_k))\\
    \end{bmatrix},
\end{equation}
where $\beta_k$ is the power attenuation owing to the distance between the $k$-th device and the AP, $\Delta$ is the normalized inter-antenna spacing, and $\phi_k\in[0,2\pi]$ is the azimuth angle relative to the boresight of the ULA of the AP. Meanwhile, the random NLoS component is distributed as
\begin{equation}
    \textbf{h}_k^{\text{nlos}}\sim\mathcal{CN}(\textbf{0},\textbf{R}_k).
\end{equation}
Note that
\begin{equation}
    \textbf{h}_k\sim\mathcal{CN}\left(\sqrt{\dfrac{\kappa}{1+\kappa}}\textbf{h}_k^{\text{los}},\dfrac{\textbf{R}_k}{\kappa+1}\right),
\end{equation}
where $\textbf{R}_k\in\mathbb{C}^{M\times M}$ is the positive semi-definite covariance matrix describing the spatial correlation of the NLoS components.

\par The spatial covariance matrices can be (approximately) modeled using the Gaussian local scattering model \cite[Sec. 2.6]{bjornson2017}. Specifically, the $s$-th row, $m$-th column element of the correlation matrix is
\begin{equation}
\begin{split}
    [\textbf{R}_k]_{s,m}=\dfrac{\beta_k}{N}\sum_{n=1}^N\exp[j\pi(s-m)\sin(\psi_{k,n})] \\
    \times \exp\left\{-\dfrac{\sigma_\phi^2}{2}[\pi(s-m)\cos(\psi_{k,n})]^2 \right\},
\end{split}
\end{equation}
where
$N$ is the number of scattering clusters, $\psi_{k,n}$ is the nominal Angle of Arrival (AoA) for the $n$-th cluster, and $\sigma_\psi$ is the Angular Standard Deviation (ASD).


\par The estimated channel vector of the $k$-th device, $\Hat{\textbf{h}}_k\in\mathbb{C}^{M\times1}$, can be modeled as the sum of the true channel vector plus a random error vector as \cite{wang2012,eraslan2013}
\begin{equation}
    \hat{\textbf{h}}_k=\textbf{h}_k+\Tilde{\textbf{h}}_k,
\end{equation}
where $\Tilde{\textbf{h}}_k\sim\mathcal{CN}(\textbf{0},\sigma_{\text{csi}}^2\textbf{I})$ is the vector of channel estimation errors. Note that the true channel realizations and the channel estimation errors are uncorrelated.

\par The parameter $\sigma_{\text{csi}}^2$ indicates the quality of the channel estimates. Let $\rho$ denote the transmit signal-to-noise ratio. Assuming orthogonal pilot sequences during the uplink data transmission~phase, it can be modeled as a decreasing function of $\rho$~\cite{wang2012,eraslan2013}
\begin{equation}
    \sigma_{\text{csi}}^2=\dfrac{1}{K\rho}.
\end{equation}

\subsection{Signal Model}

\par The matrix $\textbf{H}\in\mathbb{C}^{M\times K}$ containing the channel vectors of the $K$ devices transmitting their data to the AP can be written~as
\begin{equation}
    \textbf{H}=[\textbf{h}_1,\textbf{h}_2,\ldots,\textbf{h}_K].
\end{equation}
Then, the $M\times 1$ received signal vector can be written as
\begin{equation}
    \textbf{y}=\sqrt{p}\textbf{H}\textbf{x}+\textbf{n},
\end{equation}
where $p$ is the fixed uplink transmit power that is the same for all devices, $\textbf{x}\in\mathbb{C}^{K\times 1}$ is the vector of symbols simultaneously transmitted by the $K$ devices, and $\textbf{n}\in\mathbb{C}^{M\times 1}$ is the vector of additive white Gaussian noise samples such that $\textbf{n}\sim\mathcal{CN}(\textbf{0}_{M\times1},\sigma^2_n\textbf{I}_M)$. Note that $\rho=p/\sigma_n^2$.

\par Let $\textbf{V}\in\mathbb{C}^{M\times K}$ be a linear detector matrix used for the joint decoding of the signals transmitted from the $K$ devices. The received signal after the linear detection operation is split in $K$ streams and given by
\begin{equation}
    \textbf{r}=\textbf{V}^H\textbf{y}=\sqrt{p}\textbf{V}^H\textbf{H}\textbf{x}+\textbf{V}^H\textbf{n}.
\end{equation}
Let $r_k$ and $x_k$ denote the $k$-th elements of $\textbf{r}$ and $\textbf{x}$, respectively. Then, the received signal corresponding to the $k$-th device can be written as
\begin{equation}
    \label{r_k}
    r_k=\underbrace{\sqrt{p}\textbf{v}_k^H\textbf{h}_kx_k}_\text{Desired signal} + \underbrace{\sqrt{p}\textbf{v}_k^H\sum_{k'\neq k}^K \textbf{h}_{k'}x_{k'}}_\text{Inter-user interference} + \underbrace{\textbf{v}_k^H\textbf{n}}_\text{Noise},
\end{equation}
where $\textbf{v}_k$ and $\textbf{h}_k$ are the $k$-th columns of the matrices $\textbf{V}$ and $\textbf{H}$, respectively. From (\ref{r_k}), the signal-to-interference-plus-noise ratio of the uplink transmission from the $k$-th device is given by
\begin{equation}
    \label{gamma_k}
    \gamma_k=\dfrac{p|\textbf{v}_k^H\textbf{h}_k|^2}{p\sum_{k'\neq k}^K |\textbf{v}_k^H\textbf{h}_{k'}|^2+\sigma^2_n\lVert\textbf{v}_k^H\rVert^2}.
\end{equation}

\par The receive combining matrix $\textbf{V}$ is computed as a function of the matrix of estimated channel vectors $\hat{\textbf{H}}\in\mathbb{C}^{M\times K}$, $\hat{\textbf{H}}=[\hat{\textbf{h}}_1,\ldots,\hat{\textbf{h}}_K]$. In this work, we compare three different linear receive combining schemes: Maximum Ratio Combining (MRC), Zero Forcing (ZF), and Minimum Mean Square Error (MMSE). For each scheme, the receive combining matrix is computed as \cite{liu2016_2}
\begin{equation}
    \label{precodingMatrices}
    \textbf{V}=
    \begin{cases}
        \hat{\textbf{H}},&\text{ for MRC,}\\
        \hat{\textbf{H}}(\hat{\textbf{H}}^H\hat{\textbf{H}})^{-1},&\text{ for ZF,}\\
        \left(\hat{\textbf{H}}\hat{\textbf{H}}^H+\dfrac{\sigma_n^2}{p}\textbf{I}_M\right)^{-1}\hat{\textbf{H}},&\text{ for MMSE.}
    \end{cases}
\end{equation}

\subsection{Performance Metrics}

\par We adopt as the performance metric the per-user mean achievable uplink Spectral Efficiency (SE). The achievable uplink SE of the $k$-th device is \cite{liu2020}
\begin{equation}
    \label{per-user-achievable-SE}
    R_k=\mathbb{E}_{\textbf{H}}\{\log_2(1+\gamma_k)\}.
\end{equation}

\par Then, the per-user mean achievable uplink SE is obtained by averaging over the achievable uplink SE of the $K$ devices,~i.e.,
\begin{equation}
    \label{mean-per-user-achievable-SE}
    \Bar{R}=\dfrac{1}{K}\sum_{k=1}^K R_k.
\end{equation}

\section{Optimal Angular Position}
\label{optimalAngularPosition}

\par In this section, we introduce the scheme utilized for the computation of the optimal rotations of the RULA. 
We assume the AP has estimates of the positions of the devices registered to the network. Based on their positions, the AP computes the optimal angular position for the RULA. Note that high precision, high speed industrial servo-motors can operate with angular speeds of few thousands of rotations per minute \cite{Constar,Yaskawa}. Thus, they can change their angular positions in a few milliseconds, allowing the AP to track network slow-to-moderate dynamics in practice.

\par Before the uplink data transmission phase, the CPU does not have any CSI. Thus, the only information that it has available for the computation of the optimal angular position of the RULA are the estimates of the positions of the devices. Thus, we adopt a location-based beamforming \cite{maiberger2010,kela2016} approach to determine the objective function, i.e., to estimate the per-user mean achievable uplink SE for each possible rotation.

\par Herein, we assume that the location of the AP is perfectly known. Based on $\Hat{\textbf{p}}_k,\;\forall k$, the CPU computes the estimates for the distances and azimuth angles between the AP and the devices, i.e. $\hat{d}_{k}$ and $\hat{\phi}_{k}$, $\forall k$. Then, it computes pseudo channel vectors assuming full-LoS propagation as
\begin{equation}
    \textbf{h}_k^{\text{pseudo}}=\sqrt{\Hat{\beta}_k}
    \begin{bmatrix}
        1\\
        \exp(-j2\pi\Delta\sin(\Hat{\phi}_k))\\
        \exp(-j4\pi\Delta\sin(\Hat{\phi}_k))\\
        \vdots\\
        \exp(-j2\pi(S-1)\Delta\sin(\Hat{\phi}_k))\\
    \end{bmatrix}.
\end{equation}
Note that the estimated large-scale fading coefficient $\Hat{\beta}_k$ is computed as a function of the estimated distances $\Hat{d}_k$, assuming a known channel model. Receive combining vectors are then computed as a function of the pseudo channel vectors according to (\ref{precodingMatrices}). Finally, the objective function is obtained by computing the per-user mean achievable SE utilizing the pseudo-channel vectors and the corresponding receive combining vectors in (\ref{gamma_k}), (\ref{per-user-achievable-SE}), and (\ref{mean-per-user-achievable-SE}). The objective function corresponds to the per-user mean achievable SE, which is predicted assuming assuming full LoS propagation, versus the rotation of the RULA.

\par
\begin{figure}[t]
    \centering
    \includegraphics[scale=0.4]{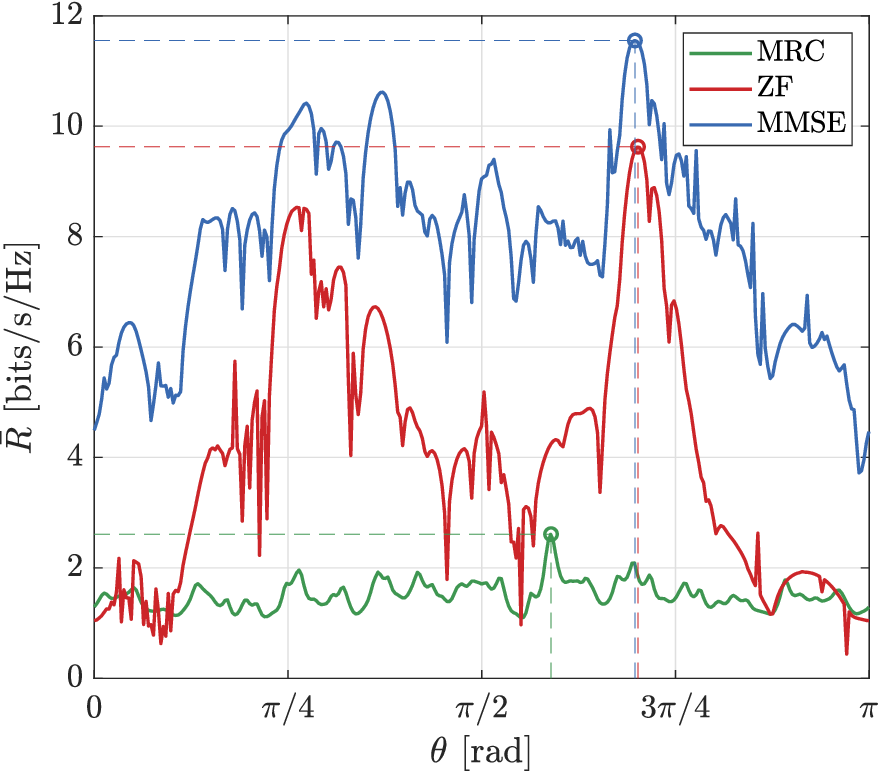}
    \caption{Predicted per-user mean achievable SE versus angular position of the RULA, for $M=16$, $K=16$ and $l=50$ m.}
    \label{objectiveFunction}
\end{figure}

 The objective function appears in Fig. \ref{objectiveFunction} for $M=16$, $K=16$ and $l=50$ m. In this figure, the circle markers show the optimal points (that is, the optimal rotation of the RULA and the corresponding value of $\Bar{R}$) for each of the three receive combining schemes considered in this work. Note that the optimal rotation of the RULA when MRC is adopted is different than the optimal rotation when ZF or MMSE are adopted. Moreover, the objective functions have several local minimum and maximum points. Thus, it is not possible to obtain the optimal rotation of the RULA using traditional optimization methods. In order to perform this task, one can resort to brute force search, or global optimization methods such as simulated annealing or particle swarm optimization \cite{engelbrecht2007}. Since we are optimizing the angular position of a single AP, i.e. our optimization problem have only one variable, we adopt brute force search in this work. Nevertheless, the other optimization methods lead to the same results.


\section{Positioning Error Model}
\label{positioningErrorModel}

\par As mentioned in Section \ref{optimalAngularPosition}, the AP utilizes the information about the positions of the devices to compute the optimal rotations of the RULA. However, the position estimates are not perfect in practical systems. In this section, we propose a mathematical model for the positioning error.

\par Since all the devices are positioned at the same height $h_{\text{device}}$, the imperfect positioning impairment refers to the uncertainty on the position of the devices only on the $(x,y)$ axes. Let $\hat{\textbf{p}}_k=(\hat{x}_k,\hat{y}_k)$ denote the estimated position of the $k$-th device. The positioning error vector associated to the $k$-th device becomes
\begin{equation}
   \textbf{e}_k=\textbf{p}_k-\hat{\textbf{p}}_k=(e_{x,k},e_{y,k}),
\end{equation}
where $e_{x,k}=x_k-\hat{x}_k$ and $e_{y,k}=y_k-\hat{y}_k$ are the $x$ and $y$ components of the positioning error vector, respectively.


\par We assume the positioning error follows a bivariate Gaussian distribution with mean $\bm{\mu}=[\textbf{0}\;\textbf{0}]^T$ and covariance matrix $\bm{\Sigma}=\sigma_e^2\textbf{I}_2$ \cite{zhu2022}. Thus, $x$ and $y$ components of the positioning error vector follow a Gaussian distribution:
\begin{equation}
    e_{x,k},e_{y,k}\sim\mathcal{N}(0,\sigma_e^2).
\end{equation}

\begin{table}[t]
    \centering
    \caption{Simulation parameters.}
    \begin{tabular}{l l l}
        \toprule
        \textbf{Parameter} & \textbf{Symbol} & \textbf{Value}\\  
        \midrule
        Number of antenna elements & $M$ & $[4,16]$\\
        Number of active devices & $K$ & $[2,16]$\\
        Length of the side of the square area & $l$ & 100 m\\
        Uplink transmission power & $p$ & 100 mW\\
        PSD of the noise & $N_0$ & $4\times10^{-21}$ W/Hz\\
        Signal bandwidth & $B$ & 20 MHz\\
        Noise figure & $N_F$ & 9 dB\\
        Height of the APs & $h_{\text{AP}}$ & 12 m\\
        Height of the UEs & $h_{\text{UE}}$ & 1.5 m\\
        Carrier frequency & $f_c$ & 3.5 GHz\\
        Variance of the positioning error & $\sigma_e^2$ & $[-20,20]$ dB\\
        Rician factor & $\kappa$ & $[-10,30]$ dB\\
        \bottomrule
    \end{tabular}    
    \label{tableParameters}
\end{table}

\section{Numerical Results}
\label{numericalResults}

\par In this section, we present Monte Carlo simulation results that reveal the performance improvements obtained with the proposed RULA when compared to a static ULA.

\begin{figure*}
    \centering
    \begin{minipage}[t]{0.475\textwidth}
        \centering
        \includegraphics[scale=0.5]{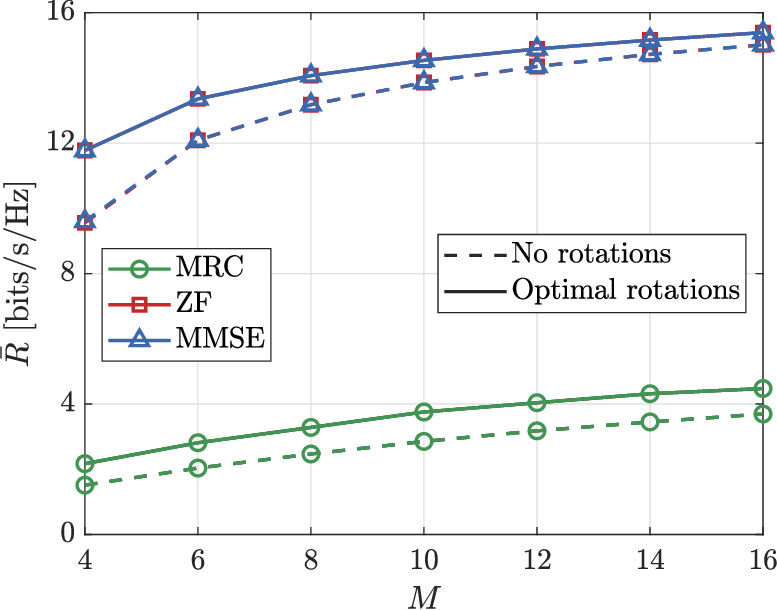}
        \caption{Per-user mean achievable SE versus number of antennas,~for $K=4$, $\kappa=10$ dB and $\sigma_e^2=-10$~dB.}
        \label{plot_numAntennas}
        \vspace{0.3cm}
        \includegraphics[scale=0.5]{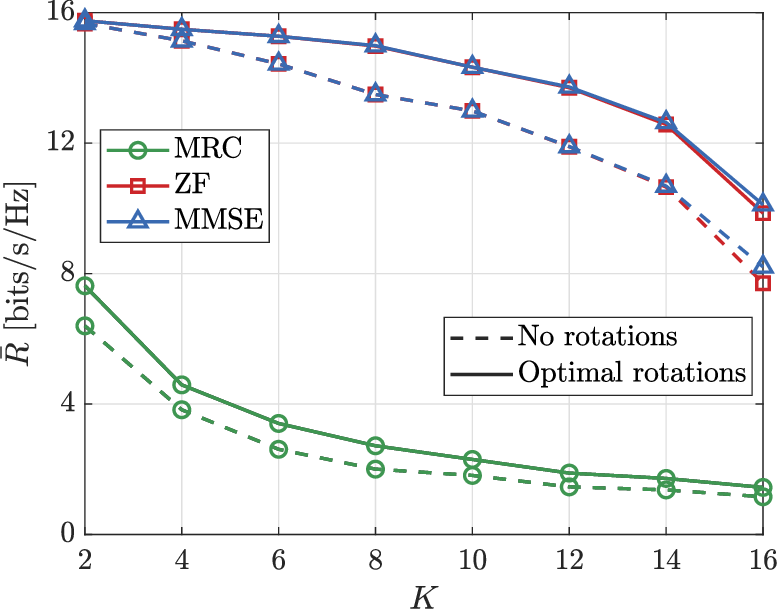}
        \caption{Per-user mean achievable SE versus number of active~devices, for $M=16$, $\kappa=10$ dB and and $\sigma_e^2=-10$~dB.}
        \label{plot_numDevices}
    \end{minipage}
    \hspace{0.4cm}
    \begin{minipage}[t]{0.475\textwidth}
        \centering
        \includegraphics[scale=0.5]{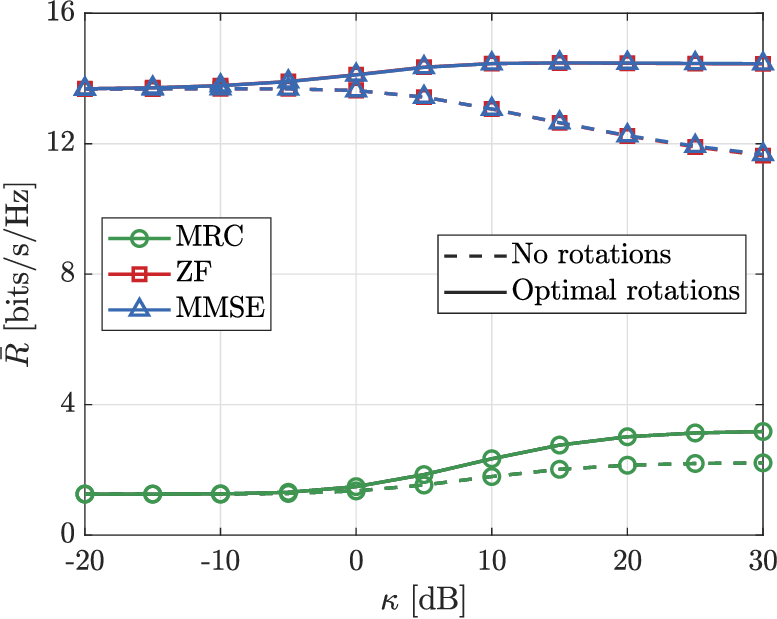}
        \caption{Per-user mean achievable SE versus Rician factor, for $M=16$, $K=10$ and $\sigma_e^2=-10$~dB.}
        \label{plot_Kappa}
        \vspace{0.3cm}
        \includegraphics[scale=0.5]{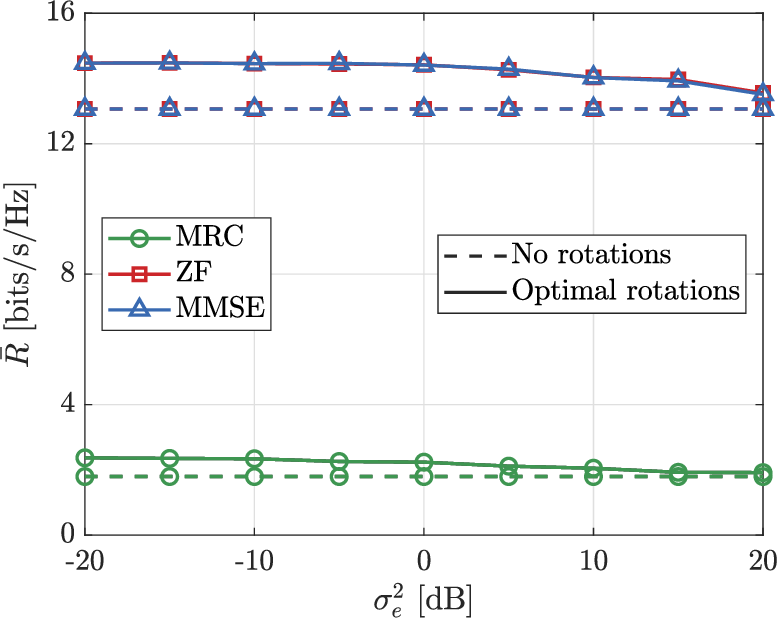}
        \caption{Per-user mean achievable SE versus variance of the positioning error, for $M=16$, $K=10$ and $\kappa=10$~dB.}
        \label{plot_posError}
    \end{minipage}
\end{figure*}

\subsection{Simulation Parameters}

\par The power attenuation due to the distance (in dB) is modeled using the log-distance path loss model as
\begin{equation}
    \beta_{k}=-L_0-10\eta\log_{10}\left(\dfrac{d_{k}}{d_0}\right),
\end{equation}
where $d_0$ is the reference distance in meters, $L_0$ is the attenuation owing to the distance at the reference distance (in dB), $\eta$ is the path loss exponent and $d_{k}$ is the distance between the $k$-th device and the AP in meters.

\par The attenuation at the reference distance is calculated using the Friis free-space path loss model and given by
\begin{equation}
    L_0=20\log_{10}\left(\dfrac{4\pi d_0}{\lambda}\right),
\end{equation}
where $\lambda=c/f_c$ is the wavelength in meters, $c$ is the speed of light and $f_c$ is the carrier frequency.

\par Unless stated otherwise, the values of the simulation parameters are listed on Table \ref{tableParameters}. Considering the selected values of $M$ and $h_{\text{AP}}$, the communication links between the AP and any device experience far-field propagation conditions. The noise power (in Watts) is given by $\sigma^2_n=N_0BN_F$, where $N_0$ is the power spectral density of the thermal noise in W/Hz, $B$ is the signal bandwidth in Hz, and $N_F$ is the noise figure at the receivers. For the computation of the correlation matrices $\textbf{R}_k,\;\forall k$, we consider $N=6$ scattering clusters, $\psi_{k,n}\sim\mathcal{U}[\phi_k-40\degree,\phi_k+40\degree]$, and $\sigma_{\psi}=5\degree$.

\subsection{Simulation Results and Discussions}

\par The positions of the devices are uniformly randomly generated in the coverage area, i.e., $x_k,y_k\sim\mathcal{U}[0,l]$. We generate average performance results for networks of $K$ devices by averaging the per-user mean achievable SE over multiple network realizations. In other words, the numerical results correspond to the expected SE performance gains for networks of $K$ devices in the considered setup. For each network realization, the achievable SE of the $K$ devices is obtained by averaging over several channel realizations, i.e. distinct realizations of the channel matrix $\textbf{H}$. For each set of parameters, we compare the performance achieved with the three linear receive combining schemes studied in this work (MRC, ZF, and MMSE). We also compare our setup where the AP is equipped with a RULA with the traditional setup where the AP is equipped with a static ULA.

\par Fig. \ref{plot_numAntennas} shows the per-user mean achievable SE versus the number of antenna elements at the AP, for $K=4$, $\kappa=10$ dB, and $\sigma_e^2=-10$ dB. We observe that SE increases with $M$ due to the enhanced beamforming gain and the spatial multiplexing capabilities of the AP. The performance of ZF matches that of MMSE, and MRC presents a poor performance compared to its counterparts because it cannot mitigate the inter-user interference. We also note that the optimal rotation of the AP improves the performance for all values of $M$. However, in the case of ZF or MMSE, the performace improvement diminishes as the number of antenna elements grows large.

\par Fig. \ref{plot_numDevices} shows the per-user mean achievable SE versus the number of active devices in each time slot, for $M=16$, $\kappa=10$ dB, and $\sigma_e^2=-10$ dB. Now we observe that the achievable SE decreases with $K$, as the inter-user interference seen by each device increases. The performance of ZF matches that of MMSE for $K<14$, while the latter performs slightly better for $K\geq14$. We observe again that MRC performs poorly compared to ZF and MMSE. Interestingly, in the case of ZF or MMSE, the performance improvement owing to the optimal rotation of the RULA increases with $K$. For MRC, we observe the opposite effect: the achievable SE values decrease with $K$. All in all, when ZF or MMSE is adopted, the results in Figs. \ref{plot_numAntennas} and \ref{plot_numDevices} show that the performance improvement brought by the rotation capabilities of the AP are more significant as $M/K$ approaches 1.

\par In Fig. \ref{plot_Kappa}, we show the per-user mean achievable SE versus the Rician factor, considering $\sigma_e^2=-10$ dB. We first observe that the performance obtained with ZF matches that of MMSE, while the performance obtained with MRC is very poor. Moreover, in the case of an AP equipped with a static ULA and using ZF or MMSE combining, the mean per-user achievable SE decreases with $\kappa$. This happens because the correlation among the channel vectors increases with the Rician factor, which degrades the performance obtained with ZF or MMSE. We observe that the optimal rotation of the RULA improves the achievable SE obtained with any of the receive combining schemes, thus being able to compensate for the increased correlation among the channel vectors. Note also that the performance improvement increases with the Rician factor.

\par We evaluate the impact of the imperfect positioning estimates in Fig. \ref{plot_posError}, which shows the per-user mean achievable SE versus the variance of the positioning error for $\kappa=10$ dB. We compare again the performance achieved with the three different combining schemes, for the cases of an AP equipped with a static ULA or with a RULA. The dashed curves corresponding to the static ULA are constant because no location-based optimization is performed in this case. Firstly, we observe again that the best performance is obtained with ZF or MMSE. We also note that, for any of the receive combining schemes, the performance gains obtained with the optimal positions of the RULA decreases as we increase the variance of the positioning error, as expected. In the case of MRC, the performance improvement obtained with the RULA vanishes for $\sigma_e^2\geq 20$ dB. On the other hand, when ZF or MMSE are adopted, the performance gains obtained with the RULA are still very significant even when the positioning accuracy of the system is very poor. 

\section{Conclusions}
\label{conclusions}

\par In this work, we proposed the utilization of an AP equipped with RULA for indoor scenarios. The optimal rotation of the RULA is computed based on the position estimates of the served devices in order to maximize the per-user mean achievable SE. We compared the performance achieved with three different linear receive combining schemes: MRC, ZF, and MMSE. Our numerical results show that the optimal rotation of the RULA can bring substantial performance improvements in terms of the per-user mean achievable SE, when compared to the case of a static ULA, when the LoS components of the signals are strong, that is, the performance improvement grows with the Rician factor. Nevertheless, the performance gains are more significant when the number of simultaneously served devices is close to the number of antenna elements at the AP. The highest values of SE are achieved by the MMSE and ZF receivers, while the performance obtained with MRC is very poor when compared to other schemes. Moreover, ZF and MMSE receivers prove to be very robust against imperfect positioning estimates.

\section*{Acknowledgment}

\par This research was financially supported by Research Council of Finland (former Academy of Finland), 6Genesis Flagship (grant no. 346208); European union’s Horizon 2020 research and innovation programme (EU-H2020), Hexa-X-II (grant no. 101095759) project; the Finnish Foundation for Technology Promotion; and in Brazil by CNPq (305021/2021-4, 402378/2021-0) and RNP/MCTIC 6G Mobile Communications Systems (01245.010604/2020-14).

\bibliographystyle{./bibliography/IEEEtran}
\bibliography{./bibliography/main}

\begin{thebibliography}{10}
\providecommand{\url}[1]{#1}
\csname url@samestyle\endcsname
\providecommand{\newblock}{\relax}
\providecommand{\bibinfo}[2]{#2}
\providecommand{\BIBentrySTDinterwordspacing}{\spaceskip=0pt\relax}
\providecommand{\BIBentryALTinterwordstretchfactor}{4}
\providecommand{\BIBentryALTinterwordspacing}{\spaceskip=\fontdimen2\font plus
\BIBentryALTinterwordstretchfactor\fontdimen3\font minus \fontdimen4\font\relax}
\providecommand{\BIBforeignlanguage}[2]{{%
\expandafter\ifx\csname l@#1\endcsname\relax
\typeout{** WARNING: IEEEtran.bst: No hyphenation pattern has been}%
\typeout{** loaded for the language `#1'. Using the pattern for}%
\typeout{** the default language instead.}%
\else
\language=\csname l@#1\endcsname
\fi
#2}}
\providecommand{\BIBdecl}{\relax}
\BIBdecl

\bibitem{heath2018}
R.~W. Heath~Jr and A.~Lozano, \emph{{Foundations of MIMO Communication}}.\hskip 1em plus 0.5em minus 0.4em\relax Cambridge University Press, 2018.

\bibitem{zhu2023_1}
L.~Zhu, W.~Ma, and R.~Zhang, ``{Modeling and Performance Analysis for Movable Antenna Enabled Wireless Communications},'' \emph{IEEE Trans. Wireless Commun.}, pp. 1--1, 2023.

\bibitem{zhu2023_2}
------, ``{Movable Antennas for Wireless Communication: Opportunities and Challenges},'' \emph{IEEE Commun. Mag.}, vol.~62, no.~6, pp. 114--120, 2024.

\bibitem{xiao2023}
Z.~Xiao \emph{et~al.}, ``{Multiuser Communications With Movable-Antenna Base Station: Joint Antenna Positioning, Receive Combining, and Power Control},'' \emph{IEEE Trans. Wireless Commun.}, vol.~23, no.~12, pp. 19\,744--19\,759, 2024.

\bibitem{pi2023}
X.~Pi \emph{et~al.}, ``{Multiuser Communications with Movable-Antenna Base Station Via Antenna Position Optimization},'' in \emph{IEEE Globecom Workshops (GC Wkshps)}, 2023, pp. 1386--1391.

\bibitem{ma2023}
W.~Ma, L.~Zhu, and R.~Zhang, ``{MIMO Capacity Characterization for Movable Antenna Systems},'' \emph{IEEE Trans. Wireless Commun.}, pp. 1--1, 2023.

\bibitem{lozano2021}
H.~Do, N.~Lee, and A.~Lozano, ``{Reconfigurable ULAs for Line-of-Sight MIMO Transmission},'' \emph{IEEE Trans. Wireless Commun.}, vol.~20, no.~5, pp. 2933--2947, 2021.

\bibitem{onel2022}
O.~L.~A. López \emph{et~al.}, ``{CSI-Free Rotary Antenna Beamforming for Massive RF Wireless Energy Transfer},'' \emph{IEEE Internet Things J.}, vol.~9, no.~10, pp. 7375--7387, 2022.

\bibitem{lin2023}
K.~Lin \emph{et~al.}, ``{On CSI-Free Multiantenna Schemes for Massive Wireless-Powered Underground Sensor Networks},'' \emph{IEEE Internet Things J.}, vol.~10, no.~19, pp. 17\,557--17\,570, 2023.

\bibitem{ozdogan2019}
{\"O}.~{\"O}zdogan, E.~Björnson, and E.~G. Larsson, ``{Massive MIMO With Spatially Correlated Rician Fading Channels},'' \emph{IEEE Trans. Commun.}, vol.~67, no.~5, pp. 3234--3250, 2019.

\bibitem{dileep2021}
D.~Kumar \emph{et~al.}, ``{Latency-Aware Joint Transmit Beamforming and Receive Power Splitting for SWIPT Systems},'' in \emph{IEEE 32nd Annu. Int. Symp. Personal Indoor Mobile Radio Commun. (PIMRC)}, 2021, pp. 490--494.

\bibitem{bjornson2017}
E.~Bj{\"o}rnson \emph{et~al.}, ``{Massive MIMO networks: Spectral, energy, and hardware efficiency},'' \emph{Foundations and Trends{\textregistered} in Signal Processing}, vol.~11, no. 3-4, pp. 154--655, 2017.

\bibitem{wang2012}
C.~Wang, T.~C.-K. Liu, and X.~Dong, ``{Impact of Channel Estimation Error on the Performance of Amplify-and-Forward Two-Way Relaying},'' \emph{IEEE Trans. Veh. Technol.}, vol.~61, no.~3, pp. 1197--1207, 2012.

\bibitem{eraslan2013}
E.~Eraslan, B.~Daneshrad, and C.-Y. Lou, ``{Performance Indicator for MIMO MMSE Receivers in the Presence of Channel Estimation Error},'' \emph{IEEE Wireless Commun. Lett.}, vol.~2, no.~2, pp. 211--214, 2013.

\bibitem{liu2016_2}
M.~Liu \emph{et~al.}, ``{Non-Asymptotic Outage Probability of Large-Scale MU-MIMO Systems with Linear Receivers},'' in \emph{IEEE 84th Veh. Technol. Conf. (VTC-Fall)}, 2016, pp. 1--5.

\bibitem{liu2020}
P.~Liu \emph{et~al.}, ``{Spectral Efficiency Analysis of Cell-Free Massive MIMO Systems With Zero-Forcing Detector},'' \emph{IEEE Trans. Wireless Commun.}, vol.~19, no.~2, pp. 795--807, 2020.

\bibitem{Constar}
\BIBentryALTinterwordspacing
{Constar}, ``{Precision Servo Motor},'' [Accessed: Oct. 25, 2023]. [Online]. Available: \url{http://constarmotor.com/productlist/2168.html#E_2189}
\BIBentrySTDinterwordspacing

\bibitem{Yaskawa}
\BIBentryALTinterwordspacing
{Yaskawa}, ``{Sigma-5 Servo Products},'' [Accessed: Oct. 25, 2023]. [Online]. Available: \url{https://www.yaskawa.com/products/motion/sigma-5-servo-products}
\BIBentrySTDinterwordspacing

\bibitem{maiberger2010}
R.~Maiberger, D.~Ezri, and M.~Erlihson, ``{Location Based Beamforming},'' in \emph{IEEE 26th Convention Electrical Electron. Eng. Israel}, 2010, pp. 000\,184--000\,187.

\bibitem{kela2016}
P.~Kela \emph{et~al.}, ``{Location Based Beamforming in 5G Ultra-Dense Networks},'' in \emph{IEEE 84th Veh. Technol. Conf. (VTC-Fall)}, 2016, pp. 1--7.

\bibitem{engelbrecht2007}
A.~P. Engelbrecht, \emph{{Computational Intelligence: an Introduction}}.\hskip 1em plus 0.5em minus 0.4em\relax John Wiley \& Sons, 2007.

\bibitem{zhu2022}
B.~Zhu, Z.~Zhang, and J.~Cheng, ``{Outage Analysis and Beamwidth Optimization for Positioning-Assisted Beamforming},'' \emph{IEEE Commun. Lett.}, vol.~26, no.~7, pp. 1543--1547, 2022.

\end{thebibliography}

\end{document}